# Numerical evaluation of code live-load models for estimating the forces caused by actual vehicles that act on bridge substructures

Alejandro Hernández–Martínez[*], Adrián D. García–Soto and Jesús G. Valdés–Vázquez

Civil and Environmental Department, Engineering Division, Universidad de Guanajuato

Av. Juárez 77, Zona Centro, Guanajuato, Gto., México. C.P. 36000

alejandro.hernandez@ugto.mx, adgarcia@ugto.mx, valdes@ugto.mx

## ABSTRACT

The present paper assesses the efficacy of code live-load models in accurately estimating the vehicular loads transferred to bridge substructures, such as abutments, piers, and foundations. Realistic traffic vehicle data are represented using four Weigh-in-Motion databases, which provide an authentic representation of vehicle information, thus providing a realistic basis for the examination of the bridges studied. The evaluation includes various bridge models, such as single-span girder bridges and two-, three-, and four-span continuous pinned-support girder bridges. By analyzing exceedance rates, the study compares the extreme force values obtained for vehicles in the databases with those predicted by selected code live-load models. These exceedance rates are presented in spectra format, as a function of the span length. The significant variations observed in the exceedance rates highlight the need for improving existing code live-load models to achieve more accurate estimations of the forces transferred to bridge substructures. Such improvements would lead to more uniform reliability levels for any limit state, such as resistance, fatigue, serviceability, and cracking.

---

[*] Corresponding author

## 1 Introduction

The bridge design process requires different load types to be considered in order to evaluate the magnitude forces acting on structural bridge elements. One of the most important load types to consider in bridge design is vehicular live loads, *i.e.* the forces that vehicles exert on a bridge as they cross. As it is virtually impossible to predict the different types of vehicles and conditions that will be encountered throughout the life cycle of a bridge, design codes stipulate live load models (LLMs) that attempt to be representative of real traffic conditions. However, most code LLMs aim to predict the maximum force effects that act on bridge superstructures and are formulated, in most cases, based on only one-span pinned-support girder bridges and without a procedure for estimating the force magnitudes that act on substructure elements such as piers, abutments, and foundations.

One crucial aspect that is not always considered in the analysis and evaluation of bridges is that the application of code LLMs to simulate the maximum force effects exerted on a bridge superstructure does not necessarily produce the maximum force effects exerted at its supports, as shown by Williams & Hoit [1]. Currently, few studies have adequately evaluated the forces acting on bridge supports due to live-load action. Some studies, such as Huo & Zhang [2, 3, 4], have addressed this problem by focusing on how a skewed bridge modifies the transverse live-load distribution factors in order to estimate live-load reactions at bridge piers. Additionally, Erhan & Dicleli [5, 6] address the problem of estimating the forces acting on the substructures used for integral bridges, namely bridges which use abutments and monolithically-cast decks to form a rigid frame structure. Furthermore, Eamon *et al*. [7] report the transverse live-load distribution factors and shear reactions for live-load continuous connections bridges, a configuration specific to this construction system.

Williams & Hoit [1] use neural networks to estimate the forces, caused by live loads, that act on bridge piers. Similarly, Demir & Dicleli [8] apply numerical simulations to evaluate live-load effects on hammer-head piers. These are some of the few studies that have been carried out on bridges with a more conventional structure, namely bridges without skew and their slabs supported by prestressed concrete girders. Both studies used the HL-93 live-load model proposed by the American Association of State Highway and Transportation Officials (AASHTO) [9]. It should be noted that the majority of the studies conducted in this area use only one code LLM to represent the action of traffic on bridges.

The adequate evaluation of bridge support reactions enables the design of bridge substructures that prevent the occurrence of serious strength, serviceability, or fatigue problems, such as those reported by Liu *et al.* [10], which are reported to have caused the collapse of single-column pier bridges.

Given the current lack of research on the adequate evaluation of the live loads acting on bridge supports and substructure systems, the present study evaluates single-span, two-, three- and four-span continuous pinned-support girder bridge models, as shown in Figure 1. Four Weigh-in-Motion (WIM) databases are used to represent the actual traffic conditions acting upon the bridges, while the results are compared with those obtained using code LLMs from different bridge design regulations. Currently used in a multitude of engineering applications, WIM surveys represent the impact of vehicular behaviors on various structures. These applications span a wide spectrum and encompass areas such as pavement design, which has been extended to the design of the bridges, the research context to which the present study corresponds.

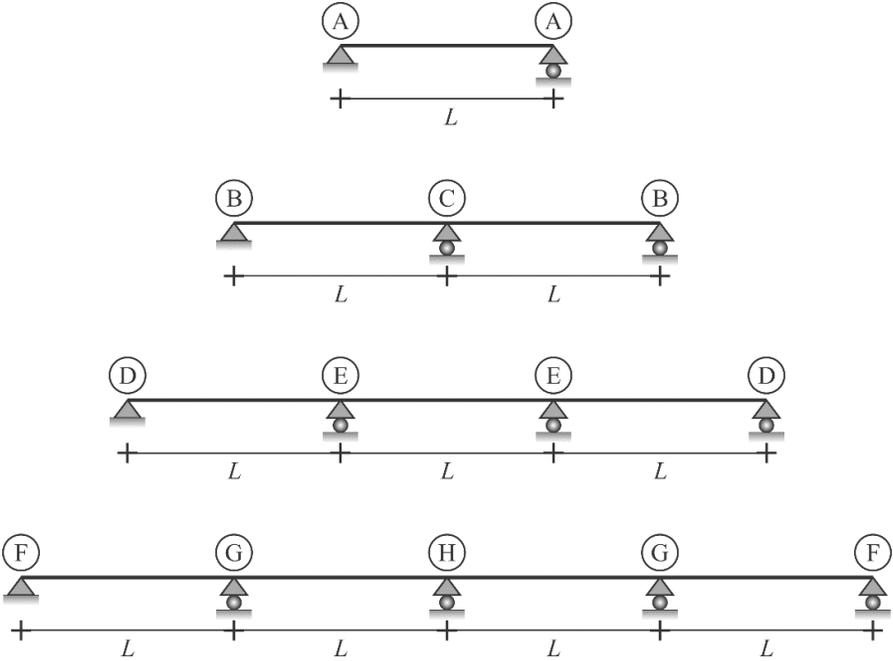

Figure 1 Bridge geometries and support designation

Given the size of the dataset to be evaluated and the disparity in the corresponding reaction force values due to the great variation in the span lengths of the bridge models studied, a direct comparison of force values becomes impractical. Therefore, the present study evaluated the reaction forces exerted on bridge supports using a conceptual adaptation of the exceedance rate, which is widely used in seismic engineering. Thus, the present paper treats exceedance rate as the number of times that the reaction force values generated by vehicles in the WIM databases exceed a reference threshold value based on a code LLM exerted on the bridge supports.

## 2 Code live-load models

Most code LLMs are proposed for estimating the most critical forces exerted by live-load effects on bridge superstructures. However, it is equally important to evaluate whether these LLMs adequately estimate the reaction forces exerted at points where the superstructure transfers forces to the bridge substructure. The substructure elements, such as piers, footings, and abutments, play a crucial role in supporting the superstructure and transmitting forces to the ground. Discrepancies between estimated and actual reaction forces can compromise the design integrity of the entire bridge, potentially leading to structural deficiencies, reduced reliability, increased risk of failure, or accelerated fatigue damage. Therefore, it is vital to evaluate the efficacy of different code LLMs in accurately estimating reaction forces in order to ensure the overall structural safety and performance of the entire bridge.

The present paper examines different LLMs obtained from various bridge design codes, including the T3-S3 (Figure 2a) and T3-S2-R4 (Figure 2b) truck models [11], although these are not applied by the current standard for bridge design in Mexico. These models are included in the present study because they represent real vehicle geometries circulating on Mexican roads and because, in contrast to most current LLMs, they do not consider a uniform lane load. The present study also considers the IMT-66.5 model (Figure 2c), which is the current LLM defined in Mexican regulations [12] and was developed based on traffic surveys conducted in the 1990s. As the AASHTO [9] standards are widely used as reference code in many countries and studies, the present study also considered the HL-93 model (Figure 2d). Additionally, in order that the complete range of the design-code LLMs used in North America, the CL-625 (Figure 2e) and CL-625-ONT (Figure 2f) models, as defined by the Canadian code, [13] are also incorporated.

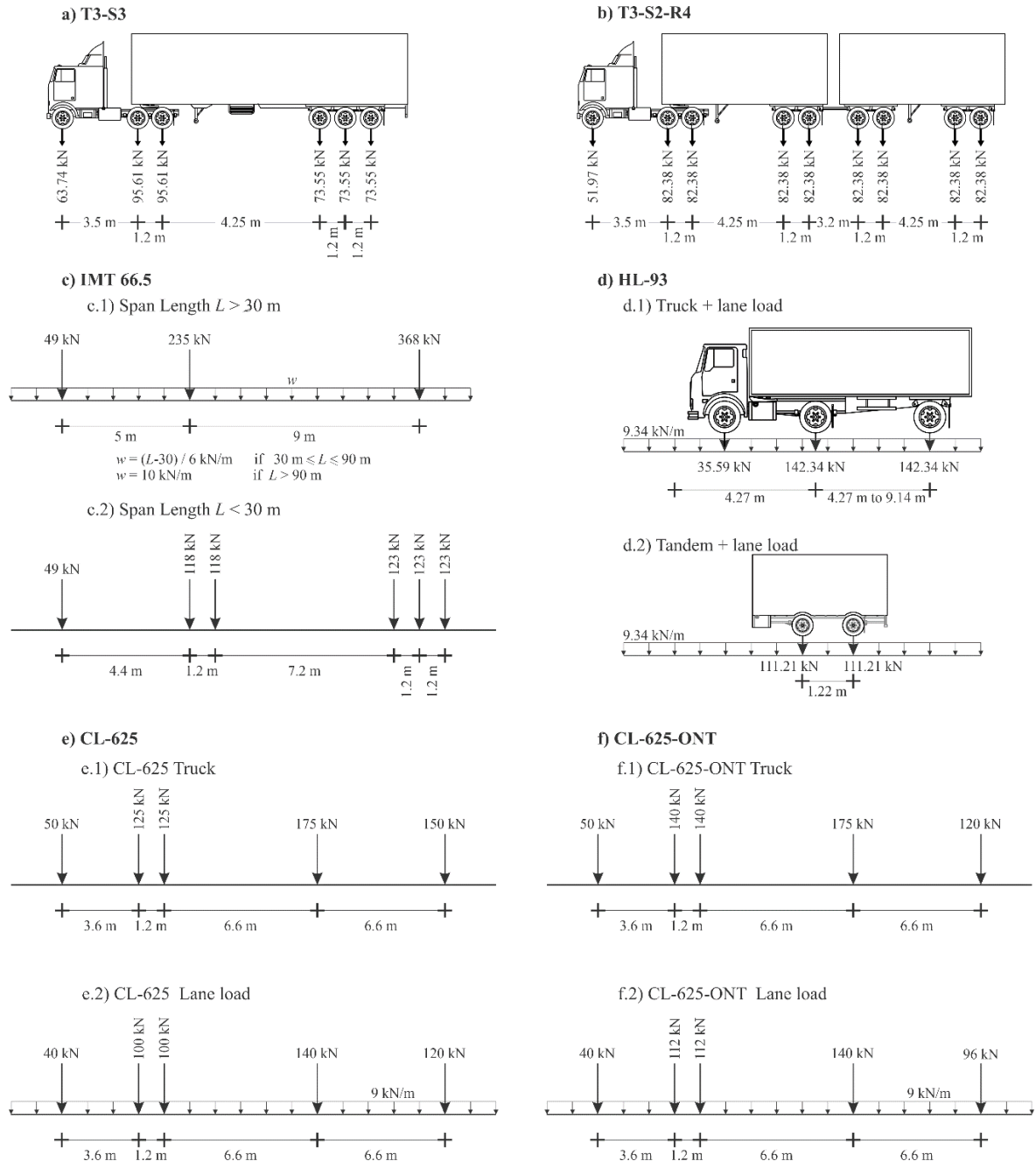

Figure 2 Code live load models

It should be noted that the present study did not include the different code LLMs mentioned above in order to rank them, namely that it did not intend to define which model was better or worse, as they present different traffic characteristics. Instead, the objective was to compare the representativeness of the force reactions obtained using code LLMs to the reactions exerted by real traffic on bridge substructures. Moreover, as no detailed bridge designs have been produced based on codes of interest, the present study evaluated neither the design philosophies nor the load and resistance design factors of such codes. The intention was not to assess whether bridges designed using these codes could withstand realistic WIM traffic data and was, instead, to evaluate the ability of the models to uniformly capture the force reactions occurring in bridge substructure elements for a wide variety of bridge types and geometries (*e.g.*, span lengths). It is expected that uniform exceedance rates would lead to more uniform reliability levels if a proper code calibration task were performed.

By comparing the force reactions obtained from code LLMs with actual traffic-induced force reactions, the present study aims to determine the inconsistencies of code LLMs in representing the actual behavior of bridge substructures. The present paper will provide valuable insights into the performance and effectiveness of code LLMs in estimating the reaction forces exerted on critical components of bridges, thus helping to enhance understanding of their suitability for structural design and the assessment of bridge substructures.

## 3    Bridge model geometries

To evaluate how code LLMs estimate support forces compared to those induced by actual traffic, the present paper proposes one-span and continuous pinned-support girder bridges with two, three, and four equal length spans, as shown in Figure 1, using the designation indicated to label the support models and results. All the bridge models examined have been computed with span lengths ranging from 1 m to 100 m, with increments of 1 m for short and medium lengths ($\leq$ 30 m) and 5 m for longer lengths ($>$ 30 m). This range of span lengths enables a comprehensive assessment of the bridge lengths commonly encountered in real-world scenarios.

By considering a wide range of span lengths and bridge configurations, the present paper assesses the precision of code LLMs in estimating support forces by comparing the results obtained to the forces generated by real vehicles. The comprehensive evaluation conducted will provide valuable insights into the reliability and effectiveness of LLMs for practical bridge design under actual traffic conditions. Ultimately, the study aims to enhance the accuracy of support force estimations, thus contributing to knowledge of the overall safety and performance of bridges under more representative real-world scenarios.

## 4   WIM databases

The present study used four WIM databases for the bridge analyses, the first of which, referred to as IP–2009, has been previously used by other studies [14, 15]. This database comprises vehicular data recorded by stationary equipment on a four-lane highway connecting the cities of Irapuato and La Piedad (Federal Highway 45 in central Mexico) from January to March 2009. The IP–2009 database contains a total of 3,832,515 vehicular records. Figure 3a shows the histogram for all vehicles in the database, based on their gross vehicular weight (GVW), while Figure 3b illustrates the histogram for vehicles with a GVW > P90(GVW), namely the histogram distribution of the heaviest 10% of vehicles in the database.

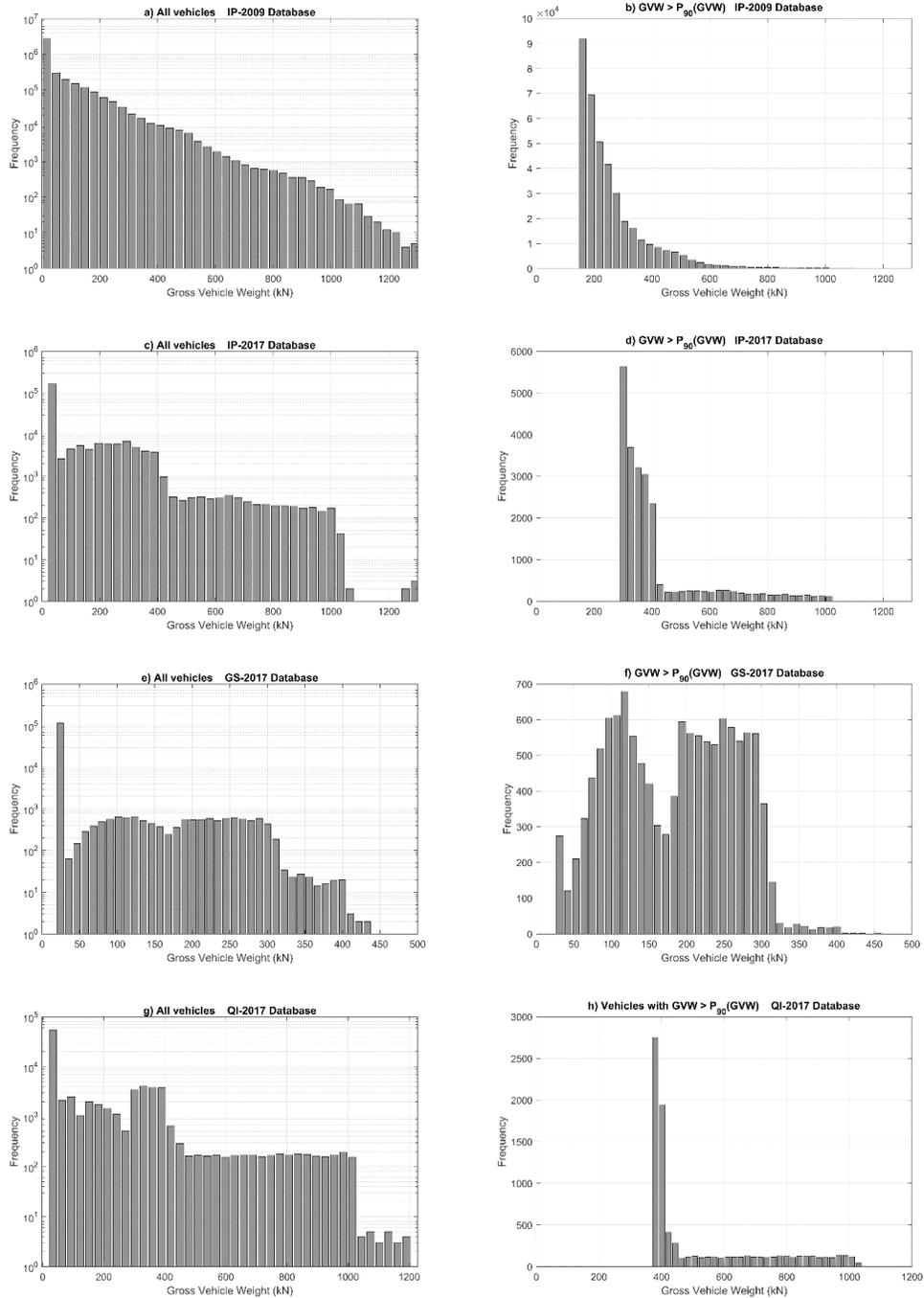

Figure 3 GVW distributions in the WIM databases

The second database used was compiled on the same highway for a duration of one week (January 24th-30th, 2017) and comprises a total of 226,489 vehicles recorded using temporary WIM recording stations. Figure 3c presents the histogram illustrating the distribution of all vehicles in the database based on their GVW. Figure 3d displays the histogram for GVW > $P_{90}$(GVW). For convenience and clarity, there WIM data will be referred to hereinafter as IP–2017.

The third database used was collected on a two-lane road connecting the cities of Guanajuato and Silao (Federal Highway 100 in central Mexico). The data was recorded for one week (January 16th-22nd, 2017) using temporary WIM equipment and comprises data corresponding to 127,218 vehicles. Figures 3e and 3f show the distribution of the vehicles recorded, based on their GVW and following the same format presented above. For convenience, this WIM database will be referred to hereinafter as GS–2017.

The fourth database used was collected on a four-lane highway (Federal Toll Road 45) connecting the cities of Querétaro and Irapuato in central Mexico. The data was recorded for a duration of one week (September 17th-23rd, 2017) using temporary WIM stations and comprises data corresponding to 86,632 vehicle records. Figures 3g and 3h show the histograms representing the distribution of vehicles based on their GVW, following the same format presented above. For ease of reference, this WIM database will be referred to as QI–2017 in the subsequent parts of this paper.

It should be noted that there is a deliberate difference in the vertical scaling in all histograms shown in Figure 3. On the left-hand side of the histograms, the vertical scale follows a logarithmic pattern, while, on the right-hand side, the vertical scale is linear. This configuration of the histograms serves the purpose of enhancing the visualization of the GVW distribution for each WIM database. In addition, Table 1 displays various measures of central tendency and dispersion for the WIM databases. Notably, it can be observed in the data presented that the GS-2017 database has the lightest average GVW. This finding is consistent with the fact that the GS-2017 database corresponds to a highway on which some of heavy configurations are not permitted. In contrast, despite comprising the least amount of vehicle data of the databases used, the QI-2017 database exhibits the heaviest average GVW. This aspect may be relevant in computing force reactions in the bridge models examined.

All the databases were collected in accordance with ASTM E318-09 [16] to ensure the reliability and consistency of the data collected, thereby maintaining the integrity of the results and analyses presented. The use of the WIM databases for analyzing various characteristics of bridges affected by live loads is not a novel approach. Several studies have previously employed this system to ascertain traffic characteristics, including the noteworthy contributions of García-Soto *et al.* [14] and Miao & Chan [17], which developed LLMs based on traffic data derived from WIM surveys, thus further exemplifying the utility of WIM data in bridge engineering research.

## 5    Software and analysis considerations

This present study used an upgraded version of AMER 2.0 software [18] to conduct all the analyses required. This software has been previously employed to both propose an LLM [14] and determine load effects in continuous span bridges [15]. The AMER software enables the computation of the extreme forces exerted on the structural elements of a bridge by each vehicle, using data obtained from the databases used by the present study (or any WIM database), as it also does for any of the LLMs considered by the present study. The analyses are performed by individually running each vehicle over the bridge models with 1 cm increments in both directions. This approach is necessary because vehicular loads are not symmetrical to their mass center, although the bridge geometries are symmetrical. One of the key features of the AMER software is its efficiency in analyzing moving loads on bridges, a capability that enables the execution of the time-consuming task of running millions of vehicles in small increments and in both directions within a reasonable time frame. This level of efficiency is difficult to achieve using commercial software, making the AMER software a valuable tool for conducting extensive bridge analysis.

Bridge structures are modeled using bar elements connected to nodes, which are considered as pinned supports, as depicted in Figure 1. The vehicle forces obtained from the WIM databases and the LLMs are applied to the bar elements, thus determining the reaction forces at each increment and enabling the extreme force reaction values (maximum and minimum) to be reported for all analyses performed.

## 6    Analysis of results

A desirable characteristic for an LLM is its ability to accurately estimate the force magnitudes exerted on all the structural elements of a bridge by vehicles throughout the service life of the bridge. As it is expected that the magnitudes and frequencies of the actual live load will vary over time, it is not surprising that the estimated force magnitudes exerted by code LLMs may be exceeded by those exerted by actual vehicles, especially the unfactored LLMs examined by the present study. However, the number of times that the estimated forces exerted by actual vehicles exceeded a reference value obtained for a code LLM should remain relatively constant. This should not only be the case across all structural elements within a given bridge but also for a wide range of bridge geometries, because code LLMs are currently applied for all bridge types without distinction. In other words, while the actual forces may surpass the estimated values obtained for code LLMs, the proportion of such occurrences should remain consistent across different structural elements of the bridge. This will facilitate the future use of a given LLM, together with calibrated load and resistance factors, to achieve more uniform reliability levels (*i.e.* probabilities of failure) across all design cases or, at least, to indicate the degree of discrepancy for a given LLM.

For a better visualization of the results, the exceedance rate spectrum (ERS) is introduced as a graphical tool to illustrate how exceedance rates for support reactions vary as a function of bridge span, for a given bridge geometry and given supports, as defined in Figure 1. The ERS is determined by computing the exceedance rate and is presented as a percentage. As discussed above, the exceedance rate is the number of times that support reactions computed for vehicles in a WIM database surpass the threshold delimited by the support reaction generated for a given LLM, which is expressed as a percentage for the ERS. Figure 4 presents the ERSs for the support reactions on a one-span bridge (identified as support A in Figure 1) and compared to each code LLM, where the x-axis represents the span length and the y-axis shows the exceedance rate. In this way, the exceedance rates, represent the proportion of instances where support reactions exceed the values generated for the LLMs, can be observed for a specific span length. Overall, the ERS offers a powerful visual representation of how well code LLMs capture the support reactions for different bridge lengths, thus enabling a comprehensive assessment of their performance in estimating the forces exerted by vehicles in the WIM databases.

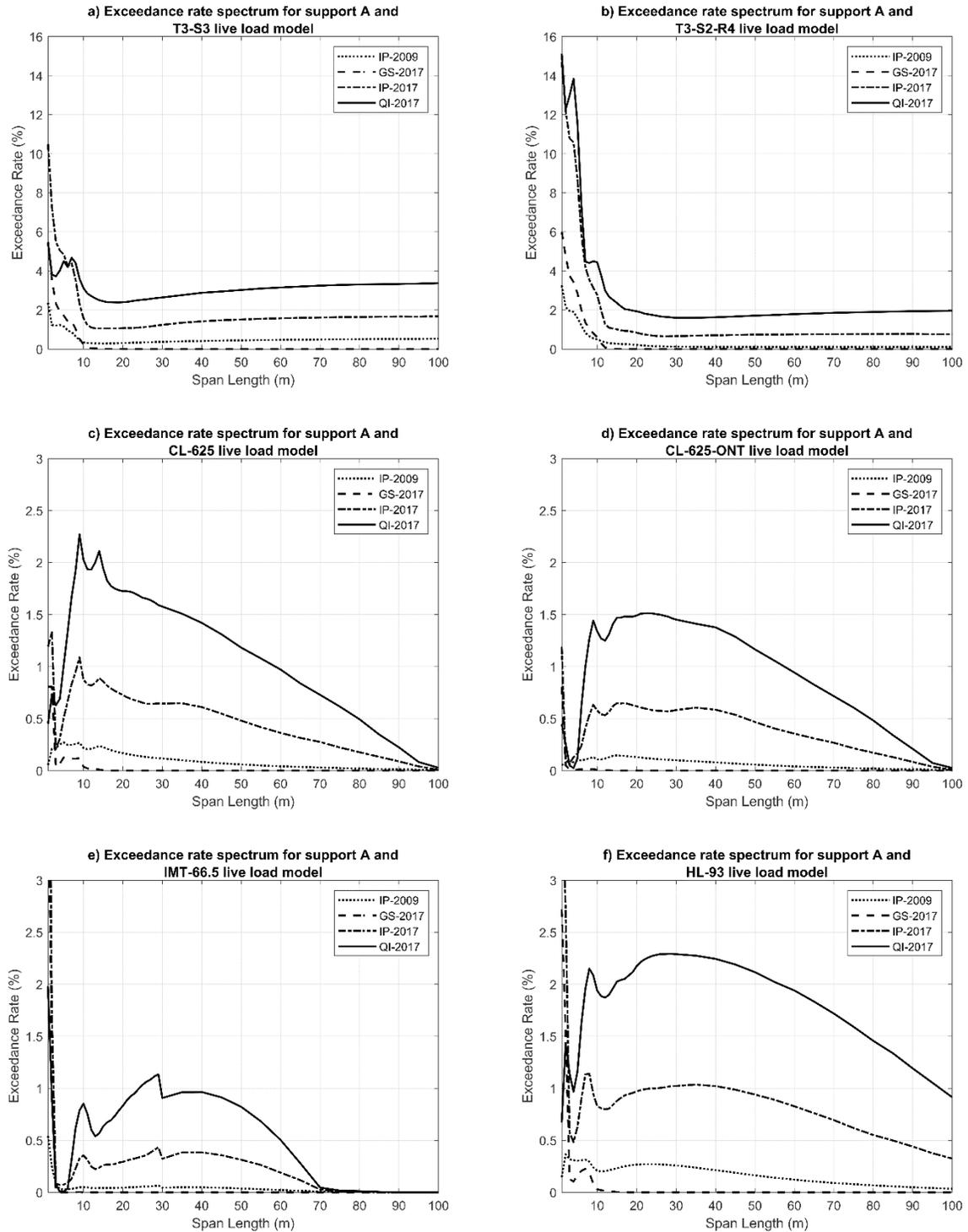

Figure 4 Exceedance rate spectra for support A

An ideal code LLM would exhibit a nearly horizontal ERS, thus indicating that the same exceedance rate should be present regardless of the bridge span length. This desired behavior would ensure a uniform level of reliability for any limit state (*e.g.* serviceability, resistance, fatigue, etc.) and would only depend on an adequately calibrated loads and the use of resistance factors together with an LLM to create the load combination for design/evaluation purposes. As can be observed in Figure 4, this does not occur for any of the LLMs considered by the present study.

For short spans, it should be noted that peak exceedance rates are inversely related to the heaviest axle weight stipulated by each LLM, namely only considering the heaviest axle in the LLM, as a heavier axle results in a lower peak exceedance rate. This inverse relation occurs because, for short spans, only one axle at a time fits over the span and, as a second axle does exert influence, the exceedance rate tends to decrease rapidly. As more axles begin to enter the span, the variation in exceedance rates tends to be more dependent on the geometry of the LLM and the vehicles in the databases, as this depends on how many and which axles fit into the bridge span, with some degree of irregularity observed in the exceedance rates.

For long spans where the LLMs fully fit into the bridge spans, two markedly different behaviors can be observed in the ERSs shown in Figure 4. For the T3-S3 and T3-S2-R4 models (Figures 4a and 4b), the exceedance rates tend to exhibit the desired constant values, becoming asymptotic as a function of the GVW for the LLMs and vehicles in the WIM databases. In contrast, for the remaining LLMs (Figures 4c to 4f), the exceedance rates tend to present a couple of peak values and then gradually decrease. These differences arise because the T3-S3 and T3-S2-R4 models do not include a uniform load in their definitions, while the other models do. It should be noted that these results are obtained by analyzing the vehicles in the databases one at a time and then comparing the reaction forces with those generated by the corresponding code LLM. To make results more compatible with models that include a uniform load, it would be necessary to consider the simultaneous presence of several vehicles obtained from the database in the lane at the same time. However, this aspect is beyond the scope of the present research, as it would require the definition of criteria for such a simultaneous presence and, due to the different natures of the code LLMs considered, this would not be a simple task and is left for future research. At this point, the significant differences shown in the ERSs presented in Figure 4 for single-span bridges should be noted, as should the high exceedance rate values obtained for short span bridges.

Approaching a span length of 30 m, the IMT-66.5 model (Figure 4e) exhibits particular behavior, which is characterized by a notable and sudden change in the exceedance rate. This abrupt change is attributed to the geometric change in the model from one that does not include a uniform load to one that does (see Figure 2c). Such abrupt changes in the exceedance rate denote inconsistencies in the LLMs, as they imply possible and considerable variations in the reliability of similar span bridges for evaluating a certain limit state.

Figure 5 illustrates de ERSs corresponding to support B for two-span bridges. Several noteworthy observations can be made based on the analysis of these ERSs. Firstly, it is evident that exceedance rate values are higher than those obtained for support A in four of the LLMs (Figures 5a to 5d). This contrast is more notable, given the peak values observed for the CL-625 and CL-625-ONT models and the trend for constant exceedance rates to present for large spans. The increase in exceedance rate values can be attributed to the fact that, in addition to the reactions observed to the vehicular loads applied, the supports themselves also experience reactive loads in the same direction as the vehicular loads (*i.e.* supports tend to uplift and reaction forces must act in the same direction as the vehicular loads applied). This behavior occurs due to the tendency of support B to uplift when vehicular loads are applied to the opposite bridge span. On the other hand, the tendency to present constant exceedance rates for long spans is also observable for the CL-625 and CL-625-ONT models (Figures 5c and 5d). This behavior is due to the geometries defined as CL-625-Truck and CL-625-ONT-Truck (Figures 2e and 2f), which do not have a uniform load in their definitions. As a result, the effect of uplift at support B is captured by these variants of the LLMs, leading to the constant exceedance rates observed. It should be noted that adequately design code normally establish the more critical cases to be used, such as placing uniform loads so that they generate the largest load effects or using reduced loads if they counteract a critical effect. [9, 13]. While thorough investigation of this is beyond the scope of the present study, it is recommended for future research. The effect of always including a uniform load becomes particularly evident when analyzing the ERSs for the IMT-66.5 and HL-93 models (Figures 5e and 5f). Due to the permanent presence of uniform loads across the entire bridge, these models are unable to predict the uplift at support B. The inability of these models to capture the uplift effect is especially noticeable for the IMT-66.5 model after a span length of 30 m, which is when the model's geometric configuration changes. Thus, obtaining exceedance rates of 100%, like those obtained for the IMT-66.5 and HL-93 models, indicates that even the lightest vehicles in the WIM databases generate an uplift effect in the supports, an aspect that LLMs considering a uniform load cannot predict.

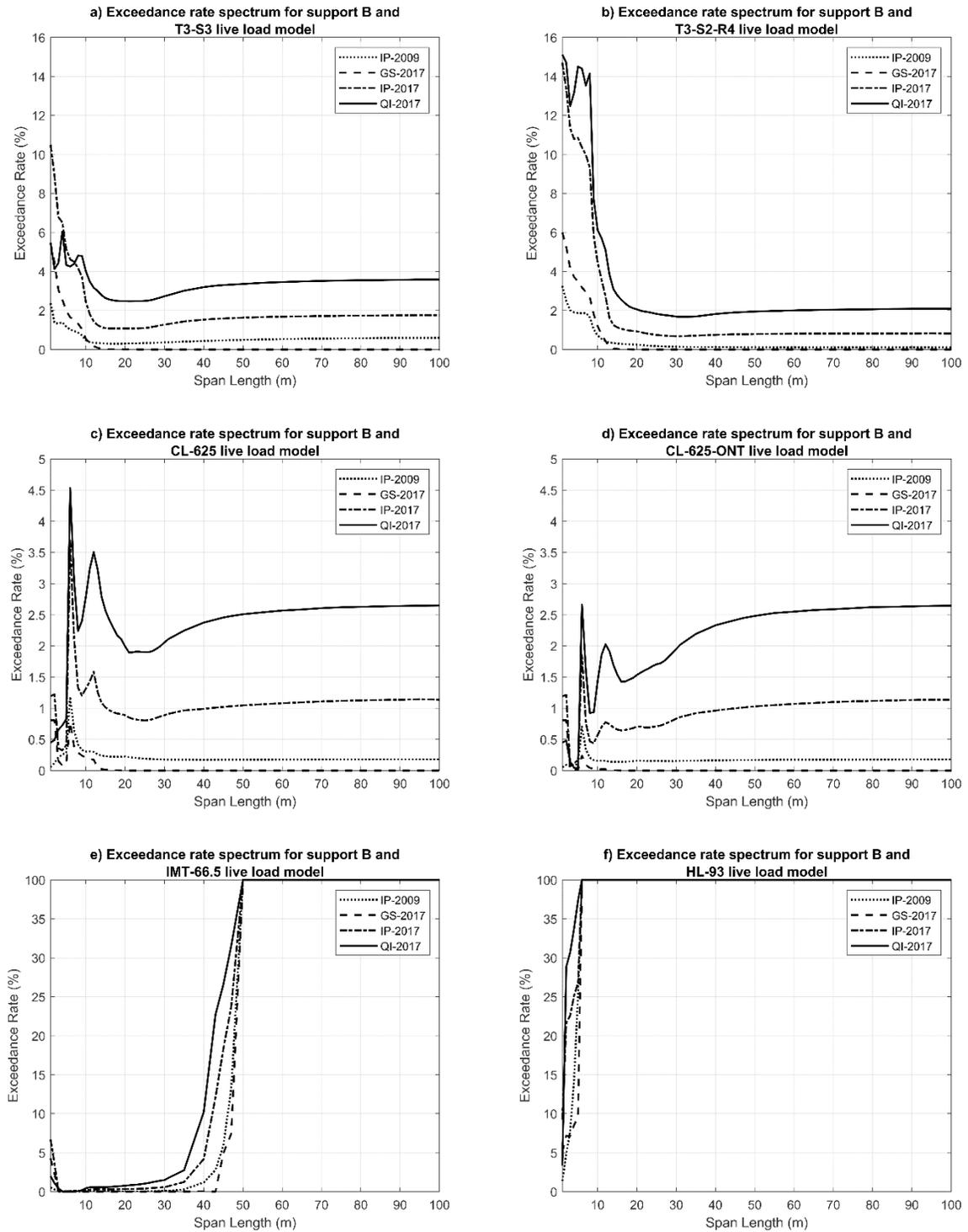

Figure 5 Exceedance rate spectra for support B

Support C, like support A, is the only one of the continuous bridge supports studied that does not exhibit reaction forces due to the tendency of the support to uplift. Figure 6 shows the ERSs corresponding to support C, from which some interesting observations can be drawn. For models T3-S3 and T3-S2-R4 (Figures 6a and 6b), the ERSs exhibit a similar behavior to those presented for supports A and B, tending to reach the same exceedance rate values for large spans. However, for the CL-625 and CL-625-ONT models (Figures 6c and 6d), there is a more pronounced decrease in the exceedance rates than in those presented in Figures 4c and 4d for support A. The ERSs show zero exceedance rates for bridge spans longer than approximately 45 m, behavior which implies that the reactions predicted by these LLMs are always greater than those generated by vehicles in the WIM databases. Furthermore, the main factor influencing this behavior is the uniform load considered by these models. Thus, for these cases, the results indicate that, for support C, the uniform load for the LLM may be too large if placed on the entire span length. While this may not be the case in the design process according to some of the codes considered, as stated above, it does point to the need to investigate this aspect in more detail in the future. Regarding the IMT-66.5 and HL-93 models, the ERSs reach values of 100%, indicating that all vehicles in the WIM databases exceed the extreme values predicted by the LLMs. This outcome occurs because the IMT-66.5 and HL-93 models are unable to predict the lower reaction forces that occur when the vehicles are about to enter or leave the bridge. If only maximum reaction force values are considered, the ERSs would present a shape more closely resembling those shown in Figures 6c and 6d but reaching zero exceedance rate values for bridge spans greater than 60 m. Although considering both the minimum and maximum values of the reaction forces for calculating exceedance rates might seem excessive, it is imperative to account for the most adverse circumstances when assessing or designing a structure. Relying solely on the maximum values does not encompass all potential unfavorable conditions.

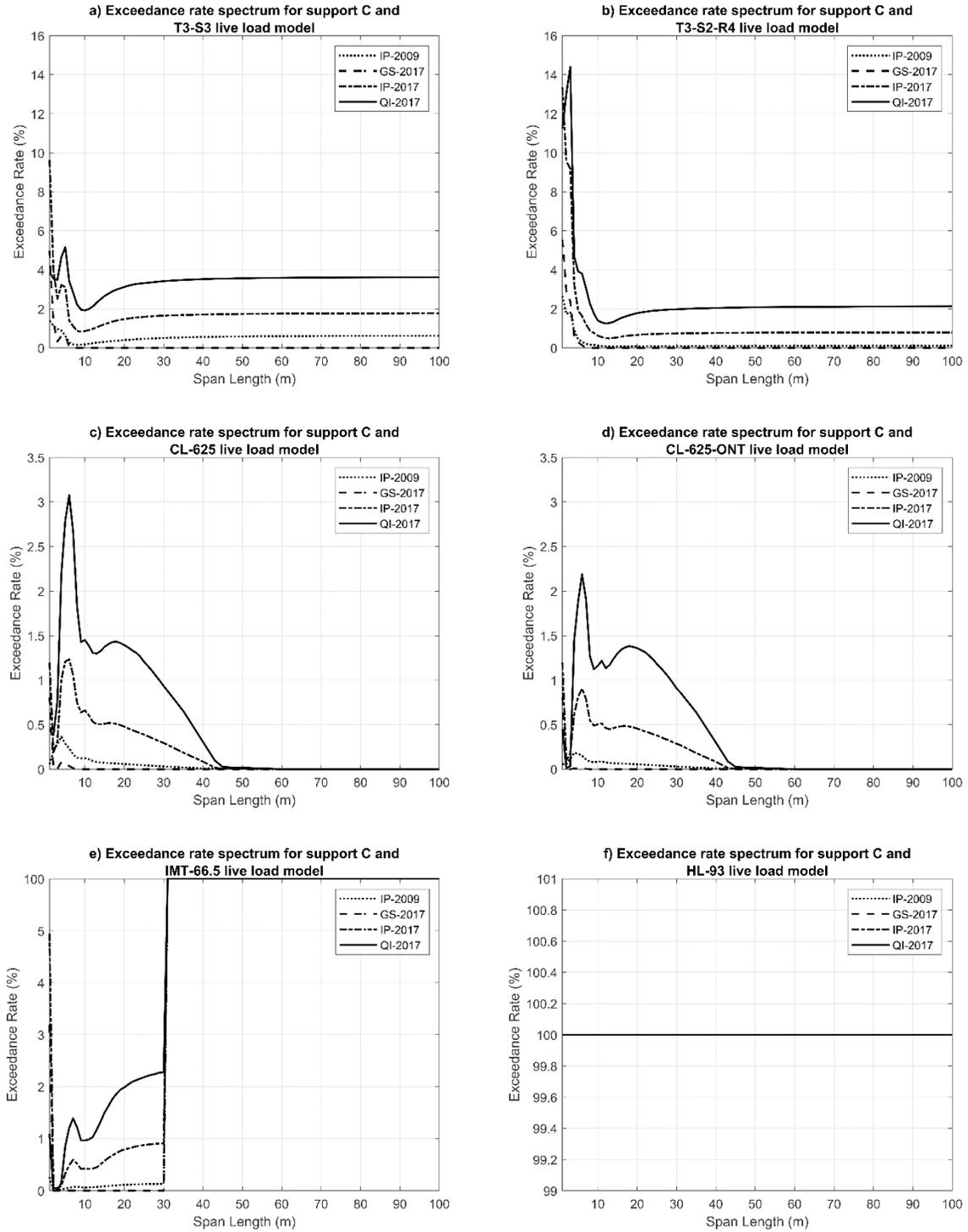

Figure 6 Exceedance rate spectra for support C

Results for three-span and four-span continuous bridges are summarized in Figures 7 and 8, which present the ERSs corresponding to supports F and H, respectively. Notably, for models T3-S3, T3-S2-R4, CL-625, and CL-625-ONT, the ERSs present similar behavior, with variations only observed for the peak exceedance rate values that are reached for each support for the small bridge spans. Subsequently, the exceedance rate values tend to remain constant as a function of GVW for each LLM, namely that the asymptotic trend value for each LLM is inversely related to their GVW. As for the IMT-66.5 and HL-93 models, the invariant presence of a uniform load throughout the bridge prevents these models from accurately predicting the tendency of the corresponding supports to uplift. As a result, the exceedance rate values for these models can reach up to 100%. These recurring outcomes stem from the intrinsic limitation of the LLMs to predict the reactive force that counters the uplift of the supports.

The inabilities, discussed above, observed for the LLMs of interest highlight the need to revise the applicability of code LLMs to accurately estimate the forces transmitted to bridge substructures, corresponding to the bridge supports in the present paper. These discrepancies are especially noticeable in short span bridges, in which sudden changes in exceedance rates are evident in the ERSs calculated. The abrupt variations in exceedance rates for short span bridges also highlight the challenges in accurately predicting support reactions for these bridge configurations.

The use of LLMs with uniform lane loads in their definitions and the absence of alternate configurations without such uniform loads, as observed in the CL-625 and CL-625-ONT models, may lead to a limitation in capturing certain unfavorable conditions, such as the potential tendency of the bridge to lift from its supports. This aspect is significant, as design considerations encompass not only the maximum force values but also the minimum values and the range within which they vary.

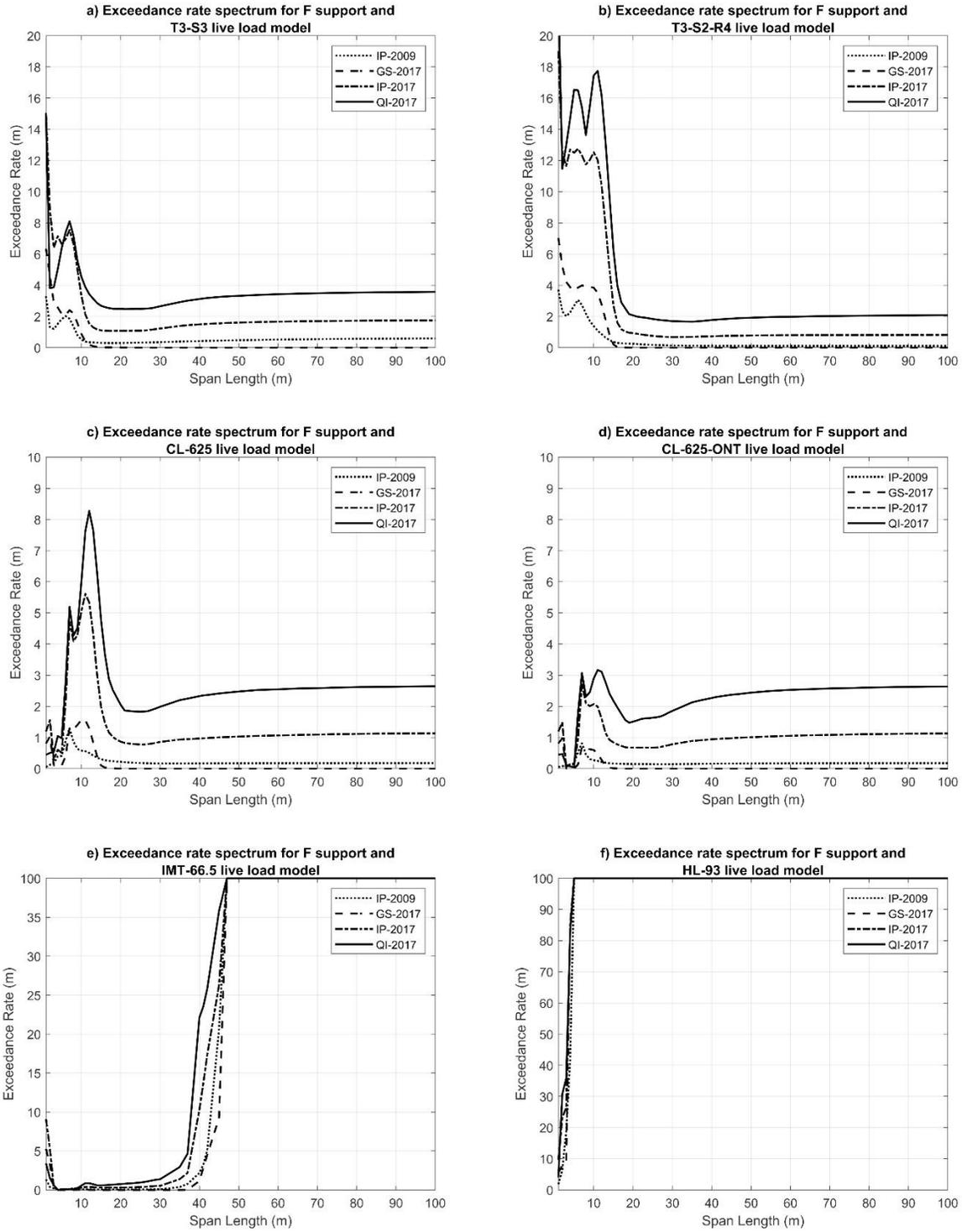

Figure 7 Exceedance rate spectra for support F

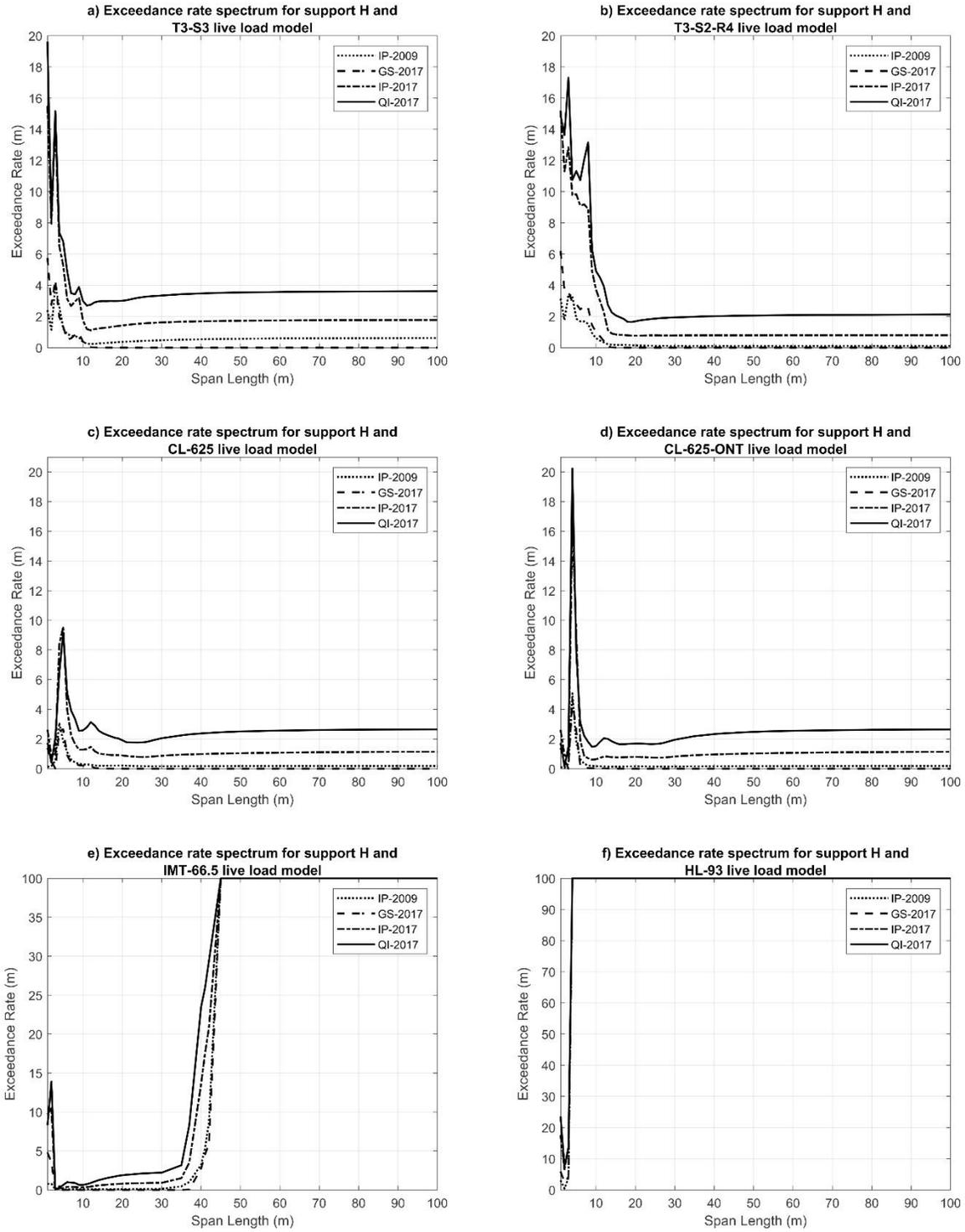

Figure 8 Exceedance rate spectra for support H

# 7 Discussion and recommendations

The LLMs were evaluated via the introduction of the ERS concept, a valuable tool that facilitates the visualization of the exceedance rate values (*i.e.* the number of times, as a percentage, that vehicles in the database surpass the value obtained from the LLMs) observed at the supports as a function of the span length. The ERS offers a comprehensive view of how well the LLMs estimate force reactions at different spans, thus helping to identify any inconsistencies or discrepancies. Thus, in an ideal scenario, the exceedance rate values observed in the spectra should exhibit constant values, meaning that exceedance rates remain practically the same regardless of the bridge span length. This ideal behavior would imply that the LLMs consistently estimate support forces with a similar reliability level for the evaluation or design of bridge elements under, perhaps, any limit state, such as resistance, serviceability, fatigue, and cracking. However, as evidenced by the results presented here, LLMs defined by code regulations struggle to replicate or approximate this ideal behavior, and, instead, present significant variations in exceedance rates, particularly for shorter bridge spans. Hence, the results obtained by the present study underscore the necessity for future research aimed at enhancing the LLMs defined in current code regulations. This enhancement process would also entail calibrating the load and resistance factors to ensure their adequate application in bridge construction practices.

Accurately estimating the forces acting on bridge supports and predicting the force magnitudes affecting the bridge superstructure and other bridge components are complex and challenging tasks. An ideal LLM must account for a wide range of factors and variables, including dynamic vehicle behavior, traffic patterns, transversal load distributions, vehicles simultaneously present in the same lane, bridge geometries, and material properties. These aspects, among many others, should be addressed in future research to improve the reliability of bridges under live-load conditions.

The specific results obtained by the present paper may vary when different databases are used for bridge analysis. However, the fundamental issue of inconsistencies among the LLMs is expected to persist across actual traffic conditions. As a result, it becomes crucial to reassess the ability of the LLMs currently employed in different codes to accurately estimate the forces transmitted to bridge substructures. This becomes evident when analyzing the results obtained for the GS-2017 database, which comprises the lightest vehicles among all the databases used by the present study. Even under such conditions, the LLMs examined here present significant fluctuations in exceedance rates for short bridge spans, which in quantity tend to be the most numerous.

Notably, the IP-2017 database shows a substantial increase in exceedance rates, in contrast to the rates obtained for the IP-2009 database on the same highway. This finding is likely not to be limited to Mexico and may be applicable worldwide. As exceedance rates increase, bridge infrastructure can become more vulnerable, leading to operational deficiencies and compromising the transportation of goods and people. Such scenario can result in significant economic repercussions, as the costs associated with maintaining a bridge in operational condition can escalate considerably. One potential avenue for mitigating these expenses is to develop more accurate LLMs that enable more realistic predictions of the forces exerted by the traffic of actual vehicles. This would aid the better preparation and maintenance of bridge infrastructure, ultimately enhancing its resilience, and reducing the economic burden associated with bridge maintenance.

In real-world conditions, the traffic scenarios encountered on bridges are diverse and dynamic, making it difficult to develop a one-size-fits-all approach. Therefore, and depending on traffic conditions, relying solely on a single LLM may not provide an accurate representation of all the scenarios needed for the comprehensive assessment or design of a bridge. Therefore, using a set of LLMs (or a set of vehicles that corresponds to the definition of the LLM) to represent various traffic conditions and bridge types may be essential to ensuring a thorough and reliable analysis. In this way, each vehicle pertaining to the set may capture specific unfavorable aspects, thus enabling a broader range of actual traffic conditions to be covered. This, in a way, already occurs with the truck and lane-load variants of the CL-625 and CL-625-ONT models (see Figures 2e and 2f). Furthermore, a potential avenue for improving LLMs in bridge engineering could be to draw inspiration from seismic engineering. In a similar vein to how synthetic seismic records can be generated to simulate seismic demands, there exists the possibility of creating "synthetic traffic". This synthetic traffic could be then used as inputs to refine and calibrate LLMs or even be used directly as LLMs, ultimately enhancing their accuracy in estimating the forces exerted on all the structural elements and predicting the overall response of bridges under more consistent live-load demands.

While achieving the level of accuracy required for synthetic traffic data might demand substantial research efforts over time, it is also possible to make simpler and more immediate improvements to LLMs. For instance, in the context of reducing exceedance rates at the supports of short span bridges, a possible improvement could involve designing LLMs in such a way that they generate greater forces along one or two of their axes. These forces could then gradually diminish as the bridge span increases, perhaps even without needing to change the GVW. However, such a hypothetical configuration, optimized to predict reactions at the supports of short span bridges, may not be suitable for accurately estimating the force magnitudes of other bridge components, such as the bridge superstructure.

One approach to accounting for the simultaneous presence of vehicles in the same lane would be akin to the concept put toward by García-Soto *et al.* [14], which involves the use of "super-vehicles". These super-vehicles are essentially a set of loads that collectively represent a convoy of vehicles traveling simultaneously. This concept enables the modeling of more representative real-world traffic scenarios in which multiple vehicles are positioned in the same lane, thus providing an alternative condition to improve the assessment of the combined effects of such traffic configurations on bridge structures. Proposing an LLM that is particularly effective for one aspect might mean that it is not as accurate for others. Achieving a balance between predicting the forces exerted on various bridge elements and maintaining overall accuracy across the entire bridge structure remains a challenge which highlights the intricacies and complexities of live-load modeling in bridge engineering.

Although the ERSs presented in this paper are computed for bridges with spans from a length of one meter upwards, they may not be directly applicable to bridges with such small dimensions. For instance, Mexican regulations define that a bridge must have spans greater than six meters, while those with shorter spans are considered drainage structures. Nevertheless, the results presented here could be applied to bridge substructures. For example, when a concrete slab deck supported on a grid of evenly spaced girders is used, the forces transferred to such elements by traffic loads and, hence, their exceedance rates would be likely to behave similarly to the results presented in the present paper for very short spans.

## 8   Conclusions

A numerical assessment of code LLMs to accurately estimate the forces generated by vehicular traffic on bridge substructures is examined by the present paper. Given the lack of studies conducted to date in this area, various bridge models, including one-, two-, three-, and four-span continuous pinned-support girder bridges were considered as study models.

Four WIM databases were employed to represent the variety of real-life vehicles recorded in central Mexican roads. The WIM databases comprise data corresponding to vehicles ranging in number from just over 86,000 vehicles for the QI-2017 database to just over 3.8 million vehicles for the IP-2009 database. An upgraded version of the AMER 2.0 software [18] was used for all the analyses conducted by the present study. The software's efficiency enabled millions of analysis cases to be performed in a reasonable timeframe, which made it possible to obtain the results presented here. The combination of diverse WIM databases and advanced software capabilities facilitates a comprehensive evaluation of code LLMs and their accuracy in predicting the forces exerted on bridge substructures under variable traffic conditions, including very heavy vehicles.

The ERS presented in the present paper provide a clear demonstration that, for short spans in all the bridge models studied here, notable high exceedance rates are observed at the supports, in contrast to that found for medium and large span bridges. This implies that the reactions estimated using any of the code LLMs examined here are insufficient for accurately predicting the forces generated by the vehicles in the WIM databases used. In fact, they consistently underestimate the actual forces exerted by the vehicles in the WIM databases. The higher exceedance rates observed for short bridges than for medium and long bridges render the substructures of these shorter bridges more vulnerable to a variety of issues. Considering that short bridges tend to be more numerous than medium and long bridges, the significance of this problem may not necessarily be in terms of its severity but more in terms of its prevalence, due to the high numbers of this kind of bridges. This highlights the importance of addressing these issues for short bridges, as collectively, they constitute a substantial concern in bridge engineering and infrastructure management.

For large bridge spans in which code LLMs fit entirely within the bridge span, it is of crucial importance to analyze the simultaneous presence of vehicles in the WIM databases to ensure that the results are more reflective of real-world conditions, an aspect to be addressed in future research. Nevertheless, the results presented in the present paper highlight an inconsistency even when solely considering a direct comparison with one vehicle at a time, which is an aspect that should not be overlooked. In these cases, it becomes evident that the GVW is the variable that appears to exert the most significant influence on the exceedance rates in which the results tend to converge.

The ERSs presented in the present paper show distinct behavior across the various supports identified in the study models (defined in Figure 1). This divergence is particularly evident in certain cases, such as the ERSs obtained for supports A and C (Figures 4 and 6, respectively), thus leading to significant disparities in peak exceedance rate values. Consequently, evaluating different limit states (resistance, fatigue, cracking, and serviceability, etc.) can yield varying reliability levels for distinct supports of the same bridge, an outcome that is not ideal. Some seasoned engineers introduce specific load conditions to mitigate the shortcomings of LLMs featuring permanent uniform loads that fail to predict support uplifting under certain conditions. This may involve loading to alternate spans or utilizing recommendations such as those appearing in the AASHTO guidelines [9], which advise the induction of negative moments in the superstructure of continuous bridges by applying a uniform load combined with two vehicular loads spaced 15 m apart. Similarly, the Canadian standards [13] prescribe a uniform load arrangement for the most critical effect. However, while these recommendations are designed to estimate extreme forces in bridge superstructures, their applicability in estimating the forces acting on substructures may not have been thoroughly evaluated. Additionally, the procedures or recommendations appearing in the design codes for this specific purpose might not be sufficiently developed. This underscores the need for further research and refinement in the domain of live-load modeling, especially when it pertains to accurately estimating force magnitudes exerted on substructure components of a bridge, such as footings, abutments, and anchors.

The noticeably higher exceedance rates observed for the IP–2017 WIM database than the IP-2009 database can be attributed to a rise in the number of vehicles and/or the loads they carry and is a significant finding. This phenomenon likely extends beyond this area of research and is indicative of broader trends in various regions of Mexico and on a global level. Therefore, enhancing LLMs is critical to the maintenance of existing infrastructure. Without more accurate LLMs, the costs associated with bridge maintenance are expected to increase considerably. Furthermore, the deterioration of bridges and especially their substructures is likely to accelerate significantly. This underscores the urgent need for improving LLMs to ensure the longevity and safety of bridge infrastructure and to manage maintenance costs effectively.

## 9 Acknowledgments

The financial support from CONAHCyT (Humanities, Sciences and Technologies Council of Mexico) Project "Problemas Nacionales 2014" No. 248165) for obtaining the IP-2017, GS-2017, and QI-2017 databases is gratefully acknowledged. The authors also like to thank the University of Guanajuato for its support to carry out this research work.

| Axles | Number of vehicles | | | | Average GVW values (kN) | | | | GVW standard deviation (kN) | | | |
|---|---|---|---|---|---|---|---|---|---|---|---|---|
| | IP-2009 | IP-2017 | GS-2017 | QI-2017 | IP-2009 | IP-2017 | GS-2017 | QI-2017 | IP-2009 | IP-2017 | GS-2017 | QI-2017 |
| All | 3,832,515 | 226,489 | 127,218 | 86,632 | 50.40 | 88.02 | 38.21 | 129.37 | 84.39 | 135.39 | 53.77 | 180.85 |
| 2 | 3,038,591 | 194,511 | 126,044 | 61,113 | 19.24 | 46.49 | 36.96 | 31.92 | 20.71 | 64.89 | 51.85 | 33.13 |
| 3 | 289,283 | 11,983 | 1,003 | 6,611 | 105.68 | 208.45 | 141.21 | 197.94 | 61.14 | 58.62 | 33.91 | 52.04 |
| 4 | 26,700 | 1,339 | 3 | 43 | 89.01 | 270.52 | 290.41 | 246.00 | 89.43 | 130.63 | 33.79 | 74.45 |
| 5 | 322,680 | 11,782 | 150 | 12,312 | 178.36 | 362.09 | 349.54 | 349.19 | 94.86 | 63.17 | 35.46 | 32.55 |
| 6 | 77,504 | 2,498 | 18 | 2,256 | 252.45 | 405.68 | 376.95 | 365.65 | 143.94 | 117.82 | 39.62 | 45.06 |
| 7 | 3,179 | 511 | | 142 | 210.28 | 481.40 | | 538.78 | 113.82 | 168.25 | | 187.33 |
| 8 | 3,041 | | | 2 | 210.09 | | | 593.55 | 167.07 | | | 297.14 |
| 9 | 71,303 | 3,865 | | 4,153 | 321.13 | 167.77 | | 658.77 | 184.16 | 220.17 | | 221.89 |
| 10 | 190 | | | | 281.64 | | | | 184.17 | | | |
| 11 | 24 | | | | 249.22 | | | | 211.40 | | | |
| 12 | 8 | | | | 154.06 | | | | 91.62 | | | |
| 13 | 6 | | | | 103.00 | | | | 89.65 | | | |

| Axles | Number of Vehicles | | | | Average heaviest axle values (kN) | | | | Heaviest axle standard deviation (kN) | | | |
|---|---|---|---|---|---|---|---|---|---|---|---|---|
| | IP-2009 | IP-2017 | GS-2017 | QI-2017 | IP-2009 | IP-2017 | GS-2017 | QI-2017 | IP-2009 | IP-2017 | GS-2017 | QI-2017 |
| All | 3,832,515 | 226,489 | 127,218 | 86,632 | 18.78 | 35.35 | 22.18 | 39.89 | 22.76 | 42.35 | 32.06 | 34.23 |
| 2 | 3,038,591 | 194,511 | 126,044 | 61,113 | 11.71 | 27.69 | 21.84 | 22.25 | 14.29 | 39.90 | 31.98 | 21.54 |
| 3 | 289,283 | 11,983 | 1,003 | 6,611 | 45.49 | 81.10 | 55.03 | 78.69 | 28.52 | 23.00 | 13.33 | 19.58 |
| 4 | 26,700 | 1,339 | 3 | 43 | 33.53 | 80.00 | 87.38 | 72.35 | 33.86 | 38.94 | 9.71 | 21.95 |
| 5 | 322,680 | 11,782 | 150 | 12,312 | 45.53 | 83.37 | 80.71 | 83.46 | 24.51 | 14.46 | 8.06 | 7.84 |
| 6 | 77,504 | 2,498 | 18 | 2,256 | 55.48 | 79.87 | 73.82 | 70.10 | 30.93 | 23.15 | 10.48 | 8.86 |
| 7 | 3,179 | 511 | | 142 | 43.40 | 77.21 | | 139.72 | 27.60 | 26.97 | | 48.37 |
| 8 | 3,041 | | | 2 | 40.00 | | | 87.67 | 32.90 | | | 44.24 |
| 9 | 71,303 | 3,865 | | 4,153 | 47.61 | 82.85 | | 88.40 | 26.00 | 28.22 | | 29.89 |
| 10 | 190 | | | | 46.16 | | | | 31.08 | | | |
| 11 | 24 | | | | 44.99 | | | | 31.51 | | | |
| 12 | 8 | | | | 26.55 | | | | 24.87 | | | |
| 13 | 6 | | | | 19.53 | | | | 12.73 | | | |

Table 1 Measures of central tendency for vehicles in the WIM databases